\documentclass[a4paper]{article}

\usepackage{spconf}
\usepackage{amsmath, amsfonts, amssymb}
\usepackage{tikz,pgfplots,multirow}
\usepackage{tabularx}
\usepackage[belowskip=-10pt,aboveskip=0pt]{caption}
\usepackage{caption}
\usepackage{graphicx}
\usepackage{algorithm}
\usepackage{algpseudocode}

 \usepackage{amsmath}
\usetikzlibrary{positioning}
\usepackage{placeins}
\usepackage{booktabs}
\def\checkmark{\tikz\fill[scale=0.4](0,.35) -- (.25,0) -- (1,.7) -- (.25,.15) -- cycle;}

\newcommand{\specialcell}[2][c]{%
  \begin{tabular}[#1]{@{}c@{}}#2\end{tabular}}
  
 \newcolumntype{Y}{>{\centering\arraybackslash}X}

\title{A comparison of streaming models and data augmentation methods \\
for robust speech recognition}
\name{Jiyeon Kim, Mehul Kumar, Dhananjaya Gowda, Abhinav Garg, Chanwoo Kim}
\address{ Samsung Research, Seoul, South Korea \\ {\small \tt {\{jstacey7.kim, mehul3.kumar, d.gowda, abhinav.garg, chanw.com\}@samsung.com}}}

\begin{document}

\maketitle
\begin{abstract}
In this paper, we present a comparative study on the robustness of two different online streaming speech recognition models: Monotonic Chunkwise Attention (MoChA) and Recurrent Neural Network-Transducer (RNN-T). We explore three recently proposed data augmentation techniques, namely, multi-conditioned training using an {\it acoustic simulator}, Vocal Tract Length Perturbation (VTLP) for speaker variability, and {\it SpecAugment}. Experimental results show that unidirectional models are in general more sensitive to noisy examples in the training set. It is observed that the final performance of the model depends on the proportion of training examples processed by data augmentation techniques. MoChA models generally perform better than RNN-T models. However, we observe that training of MoChA models seems to be more sensitive to various factors such as the characteristics of training sets and the incorporation of additional augmentations techniques. On the other hand, RNN-T models perform better than MoChA models in terms of latency, inference time, and the stability of training. Additionally, RNN-T models are generally more robust against noise and reverberation. All these advantages make RNN-T models a better choice for streaming on-device speech recognition compared to MoChA models.
 
\end{abstract}


\section{Introduction}
Recent dramatic improvement in End-to-End (E2E) Automatic Speech Recognition (ASR) systems \cite{kim2020review} has been achieved thanks to advances in deep neural network. These end-to-end speech recognition systems mainly consist of a single neural network component that performs all the equivalent tasks that are used to be performed by many discrete components in conventional speech recognition systems consisting of Acoustic Model (AM), Language Model (LM), the Pronunciation Model (PM), and so on. Connectionist Temporal Classification (CTC) \cite{a_graves_icml_2006_00}, Attention-based Encoder-Decoder (AED) model \cite{44926, gowda2020utterance}, and Recurrent Neural Network-Transducer (RNN-T) \cite{graves2012sequence, li2019improving} are often used in recent speech recognition systems. These model architectures have a simpler training pipeline with better modeling capabilities compared to conventional architectures such as DNN-HMM systems.
  
  There have been increased efforts towards ASR systems that can process streaming input in real-time, preferably on-device \cite{he2018streaming, garg2020streaming}. The transition from the server-based ASR to the on-device ASR systems reduces the cost of maintenance for service providers and more relevant for tasks where privacy, accessibility, or lower latency are required. Furthermore, an on-device streaming model can now surpass a server-side conventional model in accuracy  \cite{sainath2020streaming, garg2020hierarchical}.  

Research on various data augmentation methods has also become relevant to the need for more representative datasets for training these ASR systems. \cite{Park_2019} introduces time warping, frequency masking, and time masking for data augmentation. \cite{C_Kim_INTERSPEECH_2017_1} introduces room simulation with the different reverberation time, Signal-to-Noise Ratios (SNR), microphone, and sound source locations for far-field speech recognition. \cite{park2019specaugment} introduces SpecAugment and compares it with room simulation with different combinations. By randomly generating a warping factor, the speaker variability \cite{Kim2019, n_jaitly_icml_workshop_2013_00} can also be improved. There has also been 
effort towards alleviating the impact of noise and reverberation while training an ASR model, such as \cite{pncc_chanwoo}.
  
  Other relevant research includes how the CTC architecture has been used to predict a character for each audio frame and can be used for both the attention-based E2E model and RNN-T \cite{graves2012sequence}. CTC can be jointly attached to an attention-based E2E model \cite{hori2017advances} and can be attached with a prediction network on the RNN-T model. \cite{raffel2017online} introduces hard monotonic attention and \cite{c_chiu_iclr_2018_00} combines hard monotonic attention and soft attention.  
    In order to make one on-device model that is robust for both near-field and far-field speech data, we explore streaming models with the data augmentation method. For a relevant research, \cite{he2018streaming} uses TTS approach and \cite{moritz2020streaming} selects SpecAugment as an augmentation method. 
    
In this paper, we compare two streaming speech recognition models: RNN-T and MoChA, that are trained using the same training dataset with the same {\it on-the-fly} data augmentation methods \cite{c_kim_asru_2019_01}.

  The main contribution of this paper is that by comparing two on-line streaming models, one can gain an insight into a streaming ASR model architecture suitable for both near-field and far-field scenarios. 
  
  The rest of this paper is structured as follows: Sec. 2 describes streaming model architecture and augmentation techniques that we have used for our experiments. The experimental setup, along with details about the model composition, training strategies, hyperparameter selection, and other empirical information will be discussed in Sec. 3. Sec. 4 describes the results of our experiments, specifically the comparison between two streaming models. We conclude the paper in Sec. 5.

\section{Related research}

\begin{figure}[htb]
\begin{minipage}[b]{1.0\linewidth}
  \centering
  \centerline{\includegraphics[width=9.0cm]{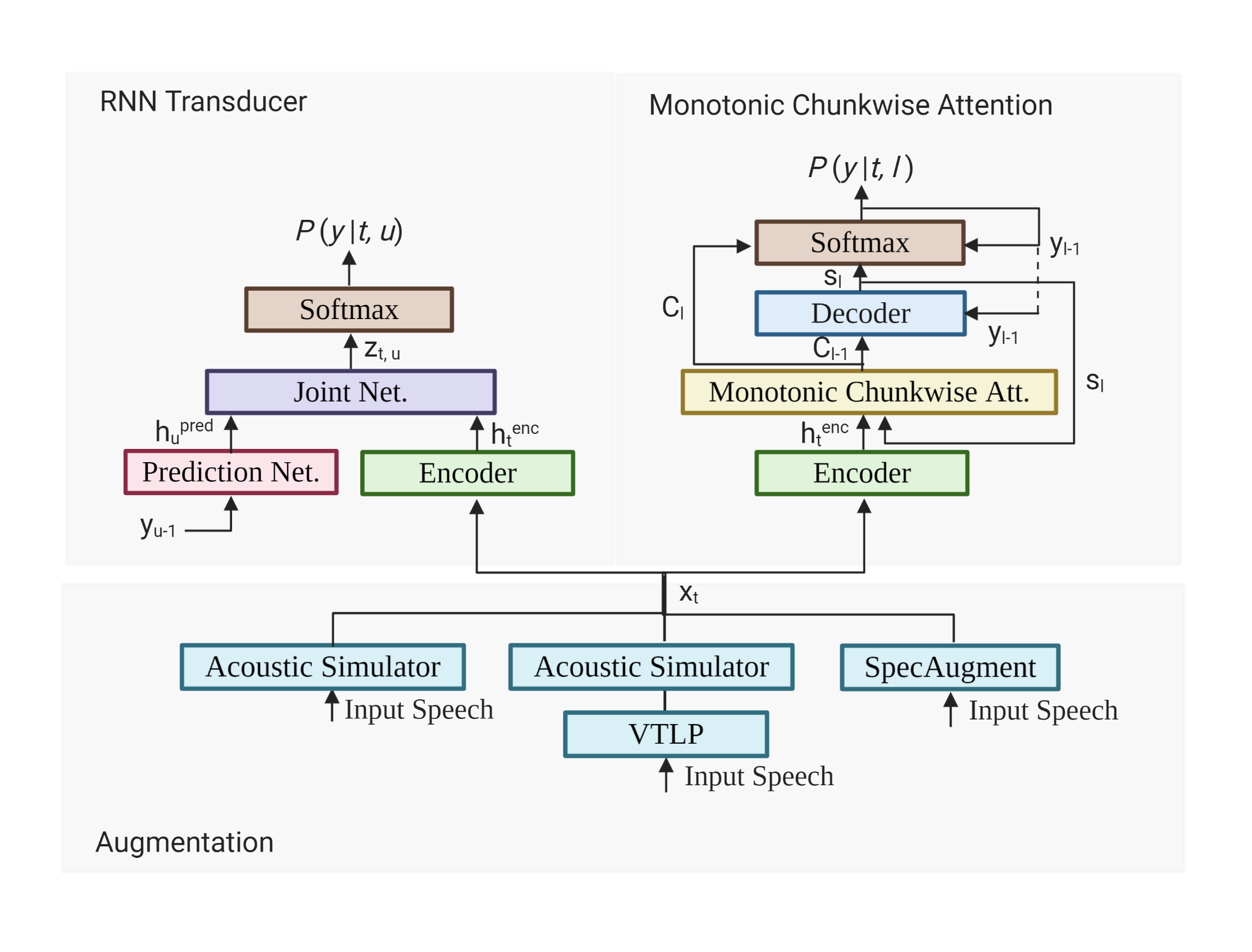}} 

\end{minipage}
\caption{Comparison of RNN-T and MoChA-based models with different data augmentation method.}
\label{fig:diagram}
\end{figure}

In this section, we give an overview of related research including streaming end-to-end speech recognition models such as MoChA and RNN-T and various data-augmentation techniques. Fig. \ref{fig:diagram} shows the structures of the RNN-T and MoChA models that were employed in our experiments.

\label{sec:format}

\subsection{Monotonic Chunkwise Attention}
\label{sssec:MoChA}
Monotonic Chunkwise Attention (MoChA) \cite{c_chiu_iclr_2018_00} is a unidirectional monotonic soft attention mechanism, derived as a hybrid of hard monotonic attention \cite{aharoni2016sequence} with local soft attention.

Input feature $x_t$ at \textit{t} timestamp, becomes an input of encoder, and encoder outputs hidden vector $h_t^{enc}$ from the input. For the attention, MoChA combines the two commonly used attention mechanisms - a probability value, calculated as the sigmoid of a global hard monotonic energy function $e_t$, is used to define the boundaries of the significant locations in the encoder sequence. Another energy function known as chunk energy $u_t$ is evaluated on a constant number, chunk size $w$, of the most recent encoder outputs. The chunk energy is then used to calculate the context vectors to be used by the decoder. With the target label $l$, decoder combines previous output $y_{l-1}$, previous context vector $c_{l-1}$ and outputs current decoder state, $s_{l}$. The model outputs a probability of label using this decoder state $s_{l}$, previous output $y_{l-1}$ and context vector $c_l$ with a softmax layer.


As an online attention mechanism, MoChA holds distinct advantages both in terms of speed and accuracy over full attention mechanisms. Also, it has a linear time complexity at the inference time and very few context vectors are calculated - proportional to the length of the output sequence, which increases the efficiency of the online model. Nevertheless, an online attention mechanism's performance suffers from the lack of complete information about the future states of the input sequence.

\begin{figure}[htb]
\begin{minipage}{0.8\textwidth}
\begin{tikzpicture}
  \node (img)  {\includegraphics[scale=0.3]{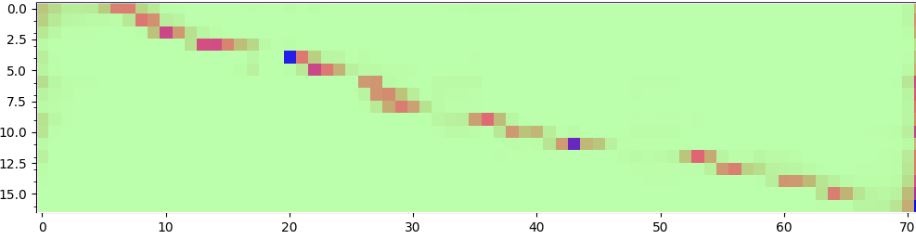}};
  \node[below=of img, node distance=0cm, yshift=1cm] {Encoder step };
  \node[left=of img, node distance=0cm, rotate=90, yshift=-0.6cm, anchor=north] {Output step};
\end{tikzpicture}
\end{minipage}%
\caption{Attention heatmap derived from MoChA, chunk size 4 }
\end{figure}

\subsection{RNN Transducer}
An RNN-T-model introduced in \cite{graves2012sequence} has been successfully employed for on-device speech recognition applications in \cite{he2018streaming}. 
As shown in Fig. \ref{fig:diagram}, an RNN-T model consists of an encoder, a prediction network, and a joint network block that combines outputs from the encoder and the prediction network. The encoder consists of six layers of LSTMs. Input feature $x_t$ at \textit{t} timestamp, becomes an input of encoder, and encoder outputs hidden vector $h_t^{enc}$ from the input feature. 
The prediction network acts as a language model and feeds the previous label output $y_{u-1}$ to predict the next label. Joint network combines those two networks, hidden vector from prediction network $h_u^{pred}$ and $h_t^{enc}$ from encoder, and outputs logits $z_{t,u}$. After passing a softmax layer, model outputs a probability of label $P(y|t,u)$. Along with CTC, and different from our attention model, RNN-T has a blank label and only non-blank output, $u^{th}$ encoder embedding output becomes an input of the prediction network. The logic of getting RNN-T logit is defined as, 
\begin{align}
z_{t,u}= f^{joint}(f^{enc}(x_t), f^{pred}(y_{u-1}))
\end{align}
where $f^{joint}$, $f^{enc}$, $f^{pred}$ are joint, encoder, and prediction logic of each network.

\subsection{Data Augmentation}
\label{sec:pagestyle}
To increase the diversity of data for training models and make a model robust, we compare streaming models with several data augmentation methods - Acoustic Simulator (AS), Vocal Tract Length Perturbation followed by the Acoustic Simulator, and SpecAugment. 

\subsubsection{Room acoustics simulation}
\label{sssec:subhead}
To enhance the robustness of models for noisy and far-field environments, we apply {\it on-the-fly} data augmentation using an {\it acoustic simulator} in \cite{C_Kim_INTERSPEECH_2017_1, c_kim_interspeech_2018_00, B_Li_INTERSPEECH_2017_1}.
  This {\it acoustic simulator}  artificially adds noise and reverberation to training utterances. This module emulates an environment for far-field speech recognition where the parameters - room dimension, microphone, and sound source location, reverberation time, SNR are randomly picked from a specific range \cite{C_Kim_INTERSPEECH_2017_1}. By doing this, we generate simulated utterances on-the-fly, thereby ensuring that the training dataset is virtually infinite in its diversity. 
   

\subsubsection{Vocal Tract Length Perturbation}
\label{sssec:VTLP}
Vocal Tract Length Perturbation (VTLP) \cite{Kim2019, n_jaitly_icml_workshop_2013_00, x_cui_taslp_2015_00} is a technique for generating a random warping factor to simulate the change in the relative length of a person's vocal tract length. This allows us to freely change voice characteristics and harmonics on the input audio. Thereby, we can bypass the restriction of having a limited number of speakers. It also helps avoid over-training biases originating from limited utterances from a type of speaker. In our experiments, we combine VTLP with AS, expecting the combined data augmentation to be effective on noise test sets. 


\subsubsection{SpecAugment}
\label{sssec:SpecAugment}
SpecAugment is introduced in \cite{Park_2019} and widely used due to its simplicity and effectiveness. SpecAugment warps and masks the features in time and frequency axis in a spectrogram. Time warping shifts a spectrogram in a time axis, time masking masks the frequencies at certain time steps. Frequency masking masks the frequencies at a certain frequency range. 

\section{Experiments}
\subsection{Experimental Setup}
For all the experiments, we employ the power-mel filterbank coefficients with a power coefficient of
 ($\cdot)^{1/15}$ \cite{Kim2019, c_kim_asru_2019_00} to extract 40 dimensions of features. We use the window size of 25 {\it ms}  with an interval of 10 {\it ms} between successive frames. As the output label, Byte Pair Encoding (BPE) is used which splits training words into 10025 units of BPE. We enable dropout rate for the encoder layer, both for MoChA and RNN-T, and 10\% label smoothing is applied to the output probability for MoChA model. For better comparison,  we use the same encoder structures with 1024 Long Short-Term Memory (LSTM) cells, 6 layers LSTM encoder and we use the beam size of 12 as a default of beam search based decoding during the inference. We trained for sufficient epochs, 13 to 15, where model converges well and performance doesn't fluctuate. All the training and testing data is 16 kHz audio. For the MoChA model, we use a chunk size of 4, and the LSTM unit size of of 1000. For better convergence, we train a model with a joint CTC and Cross-Entropy (CE) loss \cite{suyounctc} function defined as,
\begin{align}
{    
{\mathbb{L}_{Total}}  = \lambda {\mathbb{L}_{CE}}  + (1-\lambda) {\mathbb{L}_{CTC}}
      \hspace*{0.5cm} \lambda \in [0, 1]
}
\end{align}
where $\mathbb{L}_{Total}$, $\mathbb{L}_{CE}$, $\mathbb{L}_{CTC}$ are total, CE, and CTC losses respectively.

For the RNN-T model, we use a layer of 1024 units of prediction layer. We use the same training methodology and parameters of the encoder for a fair comparison. The RNN-T is trained with a CTC loss in the encoder. Combined with a prediction network, the RNN-T loss is applied to the softmax output of the joint network. As an RNN-T loss, a negative natural log of output label probability is used. 
\begin{align}
{
\mathbb{L} = {-\ln P( y | x )}
}
\end{align}

We use a linear learning rate warm-up strategy while pre-training all the encoder layers for MoChA and RNN-T models. We figure out that the model convergence is unstable as the reduction factor decreases during the pretraining stage. We trained the first layer for 0.5 epochs, followed by the addition of one LSTM layer to the encoder every 0.25 epoch. The learning rate was reset and then linearly increased when a new layer is added, as per Fig 3. We continued learning rate warm-up from 1.75 to 2.5 epochs after all the LSTM layers were added.
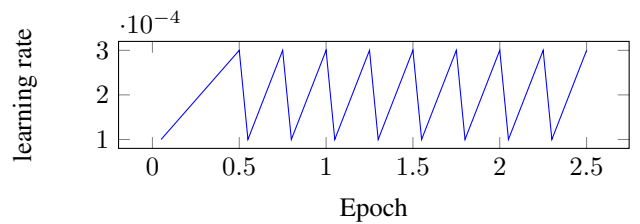
\begin{figure}[htb]
\begin{minipage}[b]{1.0\linewidth}
  \centering
\centerline{
\begin{tikzpicture}
\def\axisdefaultwidth{8.3cm}
\def\axisdefaultheight{3cm}

\begin{axis}[xlabel= Epoch,ylabel=learning rate]
\addplot[color=blue] coordinates {

(0.05,	0.0001)
(0.1,	0.00012222222222222221)
(0.15,	 0.00014444444444444444)
(0.2,	0.00016666666666666666)
(0.25,	0.00018888888888888888)
(0.3,	 0.0002111111111111111)
(0.35,	 0.00023333333333333333) 
(0.4,	0.00025555555555555553)
(0.45,	0.0002777777777777778)
(0.5,	 0.0003) 
(0.55,	0.0001)
(0.6,	0.00015000000000000001)
(0.65,	 0.00019999999999999998)
(0.7,		0.00025)
(0.75,	0.0003)
(0.8,	0.0001) 
(0.85,	0.00015000000000000001)
(0.9,	 0.00019999999999999998)
(0.95,	 	0.00025)
(1,	0.0003) 
(1.05,	0.0001)
(1.1,	0.00015000000000000001)
(1.15,	 0.00019999999999999998)
(1.2,	0.00025)
(1.25,	0.0003) 
(1.3,	0.0001) 
(1.35,	0.00015000000000000001)
(1.4,	0.00019999999999999998)
(1.45,	 0.00025) 
(1.5,	0.0003)
(1.55,	 0.0001)
(1.6,	 0.00015000000000000001)
(1.65,	 0.00019999999999999998)
(1.7,	 0.00025)
(1.75,	0.0003)
(1.8,	0.0001)
(1.85,	0.00015000000000000001)
(1.9,	 0.00019999999999999998)
(1.95,	0.00025)
(2,	 0.0003) 
(2.05,	0.0001)
(2.1,	0.00015000000000000001)
(2.15,	0.00019999999999999998)
(2.2,	0.00025)
(2.25,	0.0003)
(2.3,	0.0001)
(2.35,	 0.00015000000000000001)
(2.4,	0.00019999999999999998)
(2.45,	0.00025)
(2.5,	0.0003)
};
\end{axis}
\end{tikzpicture}
}
\end{minipage}
\caption{Learning rate warm-up schedule}
\end{figure}

\subsection{Data Augmentation}
To perform data-augmentation using an AS and VTLP, we employ an example server architecture described in \cite{c_kim_asru_2019_01}. We simulate input audio on-the-fly with 4 CPUs operating on input for each GPU. 

 In our experiments using an acoustic simulator, SNR values are sampled between 0 dB and 30 dB  from a distribution similar to that described in \cite{C_Kim_INTERSPEECH_2017_1}. Similarly, reverberation time in ($T_{60}$) is sampled from 0.0 {\it s} to 1.0 {\it s} from a distribution described in \cite{C_Kim_INTERSPEECH_2017_1}. Babble, music, and TV noises are used as noise sources. Each utterance is corrupted by one to three noise sources randomly located inside a room using an {\it acoustic simulator}. The selection probability of a noise file was dependant on the type of noise.
Room dimensions, microphone, noise,  sound source locations are randomly picked. To see the effect of data augmentation and model convergence, we enable an AS to a different percent of input data. Detailed explanation results will be shown in Sec. 4.
During experiments, a distinct effect of Acoustic Simulation on some models is observed. For the following discussion in this paper, we define $r_{\text{AS}}$ which is the  percentage of utterances processed by an acoustic simulator:

\begin{align}
{
r_{\text{AS}} = \frac{\text{\# of training utterances processed by an AS}}{\text{\# of training utterances}}\text{x}100.
}
\label{eq:r_as}
\end{align}

These `clean' audio streams still have other augmentation methods - VTLP applied to them but do not contain artificial noise or reverberation. For the VTLP configuration, we randomly choose a warping factor between 0.8 to 1.2 and oversized fft factor as 16 identical parameter setup as per \cite{Kim2019}. 



For the SpecAugment experiment, We randomly mask time and frequency on the input feature.
In a time axis, we randomly mask the number of time dimensions with a range from 1 to 20. In a frequency axis, we randomly mask up to two sections with maximum dimension 8 in 40 dimension features.

\subsection{Dataset}
\subsubsection{LibriSpeech Corpus}
LibriSpeech \cite{v_panayotov_icassp_2015_00} is a large corpus of 16kHz English speech. Each of the models presented here is trained on full 960 hours of training data available in the LibriSpeech Corpus. Both test-clean and test-other are used for performance evaluation.

\subsubsection{Test set - LibriSpeech clean with noise}
To evaluate the performance under noisy environments, we use the noisy LibriSpeech test set and the Voices test set. We synthetically add babble, music, tv noises to the clean LibriSpeech test set through AS. The same distribution of AS parameters is used during the training. 

\subsubsection{Test set - VOiCES }
VOiCES \cite{DBLP:journals/corr/abs-1804-05053} evaluation set, introduced at 2019 Interspeech Voices ASR Challenge, consists of 4600 utterances derived from the LibriSpeech dataset. 
This dataset targets acoustically challenging environments, such as noisy background, reverberation, and secondary speakers.

\section{Experimental results}
We compare MoChA and RNN-T after applying various augmentation methods -  an Acoustic  Simulator, VTLP followed by an Acoustic Simulator, and SpecAugment, in terms of the speech recognition accuracy, the number of parameters, the model size, the latency, and the inference time. For an extended comparison, we also compare them to Bi-directional LSTM with Full Attention (BFA) models and Uni-directional LSTM with Full Attention (UFA) models to provide a contrast with offline ASR models. In obtaining these results, a Language Model (LM) is not employed. For all the experimental results, we use the same beam size of 12.\\

\begin{table}[ht]
\caption{Overview of model comparison}
\centering
\begin{tabular}[t]{lcc}
\toprule
&MoChA&RNN-T\\
\midrule
Acoustic Simulator - Clean perf.&\color{green}\checkmark&\\
Acoustic Simulator - Noise perf.&&\color{green}\checkmark\\
SpecAugment perf. &\color{green}\checkmark&\\
Latency  &&\color{green}\checkmark\\
Inference time  &&\color{green}\checkmark\\
Parameter and Model size&&\color{green}\checkmark\\
\bottomrule
\end{tabular}
\end{table}%

\begin{table}[htb]
\caption{Enabling / Disabling learning rate warm-up on MoChA model, WER(\%)} 
    \centering
    \begin{minipage}{0.9\linewidth}
\begin{tabularx}{\textwidth}{c * 3 {Y}}
\hline
MoChA  & without warm-up & with warm-up\\\hline\hline
test-clean &6.41&6.18\\
 +noise &65.95&64.32\\
 test-other &18.82&18.53\\
 voices &75.31&73.51\\
 \hline
\end{tabularx}
\end{minipage}
 \end{table}
 
  \begin{table}[htp]
 \caption{Word Error Rates (WERs) (\%) of streaming Monotonic Chunkwise Attention (MoChA) and Recurrent Neural Network-Transducer (RNN-T) models trained with Vocal Tract Length Perturbation (VTLP) and an Acoustic Simulator (AS) with different $r_{\text{AS}}$ values defined in \eqref{eq:r_as}.  } 
\resizebox{\linewidth}{!}{
\vspace{4mm}
\begin{tabular}{ c|c c c c c c c }
\hline
  $r_{\text{AS}}$ &0\% &10\% & 30\% & 50\% &70\%&90\% & 100\% \\ 
\hline\hline
& \multicolumn{7}{c}{MoChA-based model} \\

test-clean &6.18&6.58&7.03&7.06&7.35&7.37&7.75 \\
 + noise &64.32&36.82&32.10&30.07&28.06&27.45&26.46\\
test-other & 18.53&18.30&18.43&18.43&17.81&18.41&18.07 \\
Voices &73.51&39.51&33.59&30.67&27.99&26.52&25.20 \\ 
 \hline
& \multicolumn{7}{c}{RNN-T-based model}   \\
test-clean & 7.86&7.83&8.11&8.55&9.55&10.42& 11.78\\
+ noise&63.03&33.97&29.81&28.32&26.70&27.01&27.92\\
test-other &20.89& 20.45&20.54&20.70&21.74&23.44&25.69 \\
Voices &71.04&36.43& 30.42&27.81&24.55&24.90&25.63 \\ 
\hline
\end{tabular}
}
\end{table}

\begin{table*}[htb]
 \renewcommand{\arraystretch}{1.3}
 \centering
 \caption{Word Error Rates (WERs) obtained with different models and different data augmentation techniques. For $r_{\text{AS}}$ in \eqref{eq:r_as}, we use a value of 70\% in experiments with an {\it acoustic simulator}.} 
\begin{minipage}{0.9\linewidth}
  \centering
  \begin{tabularx}{\textwidth}{  c * 6 {Y}}
\hline
 method & Test Set & 
\specialcell{Bi-directional\\ LSTM \\ Full Attention} & 
\specialcell{Uni-directional \\ LSTM \\ Full Attention} &
\specialcell{Uni-directional \\ LSTM \\ MoCha} & 
\specialcell{Uni-directional \\ LSTM \\ RNN-T} 
 \\\hline\hline
 \multirow{5}{*}{Baseline}&test-clean&4.38&6.86&6.18&7.86\\
 &  + noise &59.67&69.79&64.32&63.03\\
 &  test-other & 14.52&19.27&18.53&20.89\\
 &  voices & 66.82&79.43&73.51&71.04\\\cline{2-6}
 & avg. & 40.52&48.73&45.17&44.91\\\hline
 \multirow{5}{*}{AS}&
test-clean & 4.15&6.98&8.33&9.44\\
&  + noise & 18.10&
24.24&25.37&25.42\\
&test-other & 11.83&17.44&18.78&21.68\\
&voices & 17.45&24.30&25.01&23.05\\\cline{2-6}
&avg.& 13.56&19.15&20.23&20.43\\
 \hline
 \multirow{5}{*}{VTLP+AS}&
test-clean & 4.29&7.76&7.35&9.55\\
&  + noise & 17.05&
26.31&28.06&26.70\\
&test-other & 11.58&18.62&17.81&21.74\\
&voices & 15.26&25.52&27.99&24.55\\\cline{2-6}
&avg. & 12.53&20.45&21.43&21.26 \\ 

 \hline
 \multirow{5}{*}{SpecAug.}&
test-clean & 3.62&6.26&6.93&7.26\\
&   + noise & 51.98&58.22&58.43&59.06\\ 
&test-other & 11.20&16.41&17.47&18.68\\
&voices & 48.93&57.85&57.55&60.48\\\cline{2-6}
&avg. & 31.58&37.81&38.13&39.66\\\hline
\end{tabularx}
\end{minipage}
\end{table*}
 \vspace{-4mm}

\subsection{Effects of the learning rate warm-up}
 We compare performance by enabling / disabling learning rate warmup. This learning rate warmup strategy shows 0.23\% to 1.8\% absolute WER difference on our test set, especially the gap is larger on noisy test sets. The results are shown in Table 2.\\

\subsection{Effects of $r_{\text{AS}}$ for data-augmentation using an acoustic simulator}

 We define noise added LibriSpeech and Voices test set as the noisy test set. We observe a trade-off between the clean and the noisy performances with streaming models. As the percentile of AS increases, Word Error Rate (WER) on clean test sets increases while WER on noisy test sets decreases. Unlike streaming models, the BFA model trained with AS has an improvement on both the clean and noisy test set as shown in Table 4. 
 
Table 3 shows that MoChA has strength in clean speech whereas RNN-T has strength in noise speech compared to MoChA. Due to the attention mechanism of MoChA - if the probability of monotonic attention is less than 0.5, encoder embedding is not attended and the model did not output a label on some test cases. We compare its CTC performance to check the encoder performance, and figure out MoChA performs better on a relatively clean set in the BPE label unit. \\

%
%
%


\subsection{Comparison of different augmentation methods}
We compare streaming MoChA and RNN-T models with different augmentation methods. The experimental results are summarized in Table 4.  MoChA-based model generally shows better performance than RNN-T when combined with different augmentation methods especially for test-clean and test-other test sets. While comparing streaming models to non-streaming models, BFA and UFA models, the BFA model shows the best performance on all test sets as expected. Notably, the BFA model shows better performance when VTLP is added before the AS. We also conclude that applying AS is critical for the model performance on noisy test sets although it degrades the performance of a clean test set on streaming models. The performance of Specaugment degrades compared to the AS trained model on the noisy test set. \\
%




\begin{table*}[htb]
\caption{Average latency and inference time of streaming models with different beam sizes. Note that the time for obtaining the encoder output is excluded.}
 \begin{minipage}{1.0\linewidth}
\begin{tabularx}{\textwidth}{ c * 6 {Y} } 
\hline
Model & beam &1 &4 & 8 & 12 
\\\hline\hline
\multirow{2}{*}{MoChA}
& latency (ms)& 14.99&41.32&78.24&119.1
\\
&inference (s) & 1.45&1.73&1.98&2.25
\\\hline
 \multirow{2}{*}{RNN-T}
 &latency (ms)&4.38& 5.34
&6.26
&7.27
\\
&inference (s) &0.56&0.71&0.83&0.92
\\\hline

\end{tabularx}
\end{minipage}
 \end{table*}
   \vspace{-4mm}
\begin{table*}[htb]
\vspace{5mm}
\small{
\caption{Number of parameters and the size of different model architectures} 
    \centering
\begin{minipage}{1.0\linewidth}    
\begin{tabularx}{\textwidth}{ c * 5 {Y} } 
\hline
Measurement &
\specialcell{Bi-directional\\ LSTM \\ Full Attention} & 
\specialcell{Uni-directional \\ LSTM \\ Full Attention} &
\specialcell{Uni-directional \\ LSTM \\ MoCha} & 
\specialcell{Uni-directional \\ LSTM \\ RNN-T} 
\\\hline\hline
\specialcell{Number of  Parameters (million)} & 187&83&85&81\\\hline
\specialcell{Model size (MB) } & 717&318&326&313\\
 \hline
\end{tabularx}
\end{minipage}
}
 \end{table*}
   \vspace{-4mm}
\subsection{Latency and Inference Time }
Latency and inference time are measured on {\tt Nvidia Tesla P100} GPU, python version 3.5.4 through python model inference code. We use 100 utterances sampled from the LibriSpeech test-clean set and calculate average latency and inference time per sentence. Sampled utterances consist of 2346
 words, and 10427 characters. Due to the identical encoder architecture of both models, we compare latency and inference time excluding encoder computation time for a better comparison. As shown in Table 5, with a simple decoding process compared to the attention-based model \cite{li2019improving}, RNN-T shows less time on both latency and inference time. As beam size increases, the latency of the MoChA model increased drastically due to its attention mechanism.


\subsection{Parameter \& Model size}
We compare a parameter and model size of different model architectures. Parameter and model size is calculated without model compression. As expected, BFA model capacity is increased due to the bidirectional encoding. We find that parameter and model size of MoChA is larger than that of UFA and RNN-T models, mainly because of its complex computation while getting an attention weight such as monotonic attention. RNN-T consists of the least parameter and model size, along with latency and inference time.\\

\section{Conclusions}
We compare two streaming model architectures: MoChA and RNN-T in terms of inference time, latency, and speech recognition accuracies when combined with different data augmentation techniques. We observe that compared to non-streaming models, streaming models are more sensitive to the proportion of noisy examples in the training set. The final performance of the model depends on the proportion of noisy examples generated by data augmentation techniques. It is observed that
MoChA models perform slightly better than RNN-T models in terms of speech recognition accuracy especially for clean test sets while RNN-T models perform generally better in terms of latency, inference time, and the stability of parameter convergence during training. These advantages make RNN-T models a better choice for streaming on-device speech recognition compared to MoChA models.

\bibliographystyle{IEEEtran}
\bibliography{mybib,jiyeon,refs}

\begin{thebibliography}{10}
\providecommand{\url}[1]{#1}
\csname url@samestyle\endcsname
\providecommand{\newblock}{\relax}
\providecommand{\bibinfo}[2]{#2}
\providecommand{\BIBentrySTDinterwordspacing}{\spaceskip=0pt\relax}
\providecommand{\BIBentryALTinterwordstretchfactor}{4}
\providecommand{\BIBentryALTinterwordspacing}{\spaceskip=\fontdimen2\font plus
\BIBentryALTinterwordstretchfactor\fontdimen3\font minus
  \fontdimen4\font\relax}
\providecommand{\BIBforeignlanguage}[2]{{%
\expandafter\ifx\csname l@#1\endcsname\relax
\typeout{** WARNING: IEEEtran.bst: No hyphenation pattern has been}%
\typeout{** loaded for the language `#1'. Using the pattern for}%
\typeout{** the default language instead.}%
\else
\language=\csname l@#1\endcsname
\fi
#2}}
\providecommand{\BIBdecl}{\relax}
\BIBdecl

\bibitem{kim2020review}
C.~Kim, D.~Gowda, D.~Lee, J.~Kim, A.~Kumar, S.~Kim, A.~Garg, and C.~Han, ``A
  review of on-device fully neural end-to-end automatic speech recognition
  algorithms,'' in \emph{2020 54th Asilomar Conference on Signals, Systems, and
  Computers}.\hskip 1em plus 0.5em minus 0.4em\relax IEEE, 2020, pp. 277--283.

\bibitem{a_graves_icml_2006_00}
A.~Graves, S.~Fern{\'a}ndez, F.~Gomez, and J.~Schmidhuber, ``Connectionist
  temporal classification: labelling unsegmented sequence data with recurrent
  neural networks,'' in \emph{ICML}, 2006.

\bibitem{44926}
W.~Chan, N.~Jaitly, Q.~V. Le, and O.~Vinyals, ``Listen, attend and spell: A
  neural network for large vocabulary conversational speech recognition,'' in
  \emph{ICASSP}, 2016.

\bibitem{gowda2020utterance}
D.~Gowda, A.~Kumar, K.~Kim, H.~Yang, A.~Garg, S.~Singh, J.~Kim, M.~Kumar,
  S.~Jin, S.~Singh \emph{et~al.}, ``Utterance invariant training for hybrid
  two-pass end-to-end speech recognition.'' in \emph{INTERSPEECH}, 2020.

\bibitem{graves2012sequence}
A.~Graves, ``Sequence transduction with recurrent neural networks,''
  \emph{ICML}, 2012.

\bibitem{li2019improving}
J.~Li, R.~R. Zhao, H.~Hu, and Y.~Gong, ``Improving rnn transducer modeling for
  end-to-end speech recognition,'' \emph{ASRU}, 2019.

\bibitem{he2018streaming}
Y.~He, T.~N. Sainath, R.~Prabhavalkar, I.~McGraw, R.~Alvarez, D.~Zhao,
  D.~Rybach, A.~Kannan, Y.~Wu, R.~Pang, Q.~Liang, D.~Bhatia, Y.~Shangguan,
  B.~Li, G.~Pundak, K.~C. Sim, T.~Bagby, S.-Y. Chang, K.~Rao, and
  A.~Gruenstein, ``Streaming end-to-end speech recognition for mobile
  devices,'' \emph{ICASSP}, 2018.

\bibitem{garg2020streaming}
A.~Garg, G.~P. Vadisetti, D.~Gowda, S.~Jin, A.~Jayasimha, Y.~Han, J.~Kim,
  J.~Park, K.~Kim, S.~Kim \emph{et~al.}, ``Streaming on-device end-to-end asr
  system for privacy-sensitive voice-typing.'' in \emph{INTERSPEECH}, 2020.

\bibitem{sainath2020streaming}
T.~N. Sainath, Y.~He, B.~Li, A.~Narayanan, R.~Pang, A.~Bruguier \emph{et~al.},
  ``A streaming on-device end-to-end model surpassing server-side conventional
  model quality and latency,'' vol. abs/2003.12710, 2020.

\bibitem{garg2020hierarchical}
A.~Garg, A.~Gupta, D.~Gowda, S.~Singh, and C.~Kim, ``Hierarchical multi-stage
  word-to-grapheme named entity corrector for automatic speech recognition.''
  in \emph{INTERSPEECH}, 2020.

\bibitem{Park_2019}
D.~S. Park, W.~Chan, Y.~Zhang, C.-C. Chiu, B.~Zoph, E.~D. Cubuk, and Q.~V. Le,
  ``Specaugment: A simple data augmentation method for automatic speech
  recognition,'' \emph{INTERSPEECH}, 2019.

\bibitem{C_Kim_INTERSPEECH_2017_1}
\BIBentryALTinterwordspacing
{\chanwcom}, A.~Misra, K.~Chin, T.~Hughes, A.~Narayanan, T.~N. Sainath, and
  M.~Bacchiani, ``Generation of large-scale simulated utterances in virtual
  rooms to train deep-neural networks for far-field speech recognition in
  google home,'' in \emph{Proc. Interspeech 2017}, 2017, pp. 379--383.
  [Online]. Available: \url{http://dx.doi.org/10.21437/Interspeech.2017-1510}
\BIBentrySTDinterwordspacing

\bibitem{park2019specaugment}
D.~S. Park, Y.~Zhang, C.-C. Chiu, Y.~Chen, B.~Li, W.~Chan, Q.~V. Le, and Y.~Wu,
  ``Specaugment on large scale datasets,'' \emph{ArXiv}, vol. abs/1912.05533,
  2019.

\bibitem{Kim2019}
C.~Kim, M.~Shin, A.~Garg, and D.~Gowda, ``{Improved Vocal Tract Length
  Perturbation for a State-of-the-Art End-to-End Speech Recognition System},''
  in \emph{INTERSPEECH}, 2019.

\bibitem{n_jaitly_icml_workshop_2013_00}
N.~Jaitly and G.~E. Hinton, ``Vocal tract length perturbation (vtlp) improves
  speech recognition,'' in \emph{ICML}, 2013.

\bibitem{pncc_chanwoo}
C.~Kim and R.~Stern, ``Power-normalized cepstral coefficients (pncc) for robust
  speech recognition,'' \emph{IEEE/ACM Transactions on Audio, Speech, and
  Language Processing}, vol.~24, pp. 1315--1329, 2016.

\bibitem{hori2017advances}
T.~Hori, S.~Watanabe, Y.~L. Zhang, and W.~Chan, ``Advances in joint
  ctc-attention based end-to-end speech recognition with a deep cnn encoder and
  rnn-lm,'' in \emph{INTERSPEECH}, 2017.

\bibitem{raffel2017online}
C.~Raffel, M.-T. Luong, P.~J. Liu, R.~J. Weiss, and D.~Eck, ``Online and
  linear-time attention by enforcing monotonic alignments,'' in \emph{ICML},
  2017.

\bibitem{c_chiu_iclr_2018_00}
C.-C. Chiu and C.~Raffel, ``Monotonic chunkwise attention,'' in \emph{ICLR},
  2018.

\bibitem{moritz2020streaming}
N.~Moritz, T.~Hori, and J.~L. Roux, ``Streaming automatic speech recognition
  with the transformer model,'' \emph{ArXiv}, vol. abs/2001.02674, 2020.

\bibitem{c_kim_asru_2019_01}
{\chanwcom}, S.~Kim, K.~Kim, M.~Kumar, J.~Kim, K.~Lee, C.~Han, A.~Garg, E.~Kim,
  M.~Shin, S.~Singh, L.~Heck, and D.~Gowda, ``End-to-end training of a large
  vocabulary end-to-end speech recognition system,'' in \emph{2019 IEEE
  Automatic Speech Recognition and Understanding Workshop (ASRU)}, Dec. 2019,
  pp. 562--569.

\bibitem{aharoni2016sequence}
R.~Aharoni and Y.~Goldberg, ``Sequence to sequence transduction with hard
  monotonic attention,'' \emph{ArXiv}, vol. abs/1611.01487, 2016.

\bibitem{c_kim_interspeech_2018_00}
C.~Kim, E.~Variani, A.~Narayanan, and M.~Bacchiani, ``Efficient implementation
  of the room simulator for training deep neural network acoustic models,'' in
  \emph{INTERSPEECH}, 2018.

\bibitem{B_Li_INTERSPEECH_2017_1}
{{B. Li, T. Sainath, A. Narayanan, J. Caroselli, M. Bacchiani, A. Misra, I.
  Shafran, H. Sak, G. Pundak, K. Chin, K-C Sim, R. Weiss, K. Wilson, E.
  Variani, {\chanwcom}, O. Siohan, M. Weintraub, E. McDermott, R. Rose, and M.
  Shannon}}, ``{Acoustic modeling for Google Home},'' in \emph{INTERSPEECH},
  2017.

\bibitem{x_cui_taslp_2015_00}
X.~Cui, V.~Goel, and B.~Kingsbury, ``Data augmentation for deep neural network
  acoustic modeling,'' \emph{IEEE/ACM Transactions on Audio, Speech, and
  Language Processing}, vol.~23, no.~9, pp. 1469--1477, 2015.

\bibitem{c_kim_asru_2019_00}
{\chanwcom}, M.~Kumar, K.~Kim, and D.~Gowda, ``Power-law nonlinearity with
  maximally uniform distribution criterion for improved neural network training
  in automatic speech recognition,'' in \emph{2019 IEEE Automatic Speech
  Recognition and Understanding Workshop (ASRU)}, Dec. 2019, pp. 988--995.

\bibitem{suyounctc}
S.~Kim, T.~Hori, and S.~Watanabe, ``Joint ctc-attention based end-to-end speech
  recognition using multi-task learning,'' in \emph{ICASSP}, 2017.

\bibitem{v_panayotov_icassp_2015_00}
V.~Panayotov, G.~Chen, D.~Povey, and S.~Khudanpur, ``Librispeech: an asr corpus
  based on public domain audio books,'' in \emph{ICASSP}, 2015.

\bibitem{DBLP:journals/corr/abs-1804-05053}
C.~Richey, M.~A. Barrios, Z.~Armstrong, C.~Bartels, H.~Franco, M.~Graciarena,
  A.~Lawson, M.~K. Nandwana, A.~R. Stauffer, J.~van Hout, P.~Gamble,
  J.~Hetherly, C.~Stephenson, and K.~Ni, ``Voices obscured in complex
  environmental settings {(VOICES)} corpus,'' \emph{CoRR}, vol. abs/1804.05053,
  2018.

\end{thebibliography}


\end{document}